\documentclass{adonis}
\usepackage{natbib} 
\usepackage{lettrine}
\usepackage{graphicx}
\AtBeginEnvironment{quote}{\par\small} 
\usepackage{url} 
\usepackage{amsmath,amsfonts,bbm}

\title{The Actuary’s Final Word on Algorithmic Decision Making}
\author{Benjamin Recht}
\affiliation{University of California, Berkeley}
\correspondence{brecht@berkeley.edu}
\version{\today}

\runningauthor{Recht}
\runningtitle{The Actuary’s Final Word}

\abstract{Paul Meehl's foundational work \emph{Clinical versus Statistical Prediction}, provided early theoretical justification and empirical evidence of the superiority of statistical methods over clinical judgment. Despite a century of empirical evidence supporting Meehl's central thesis, from early parole prediction studies in the 1920s to modern meta-analyses, confusion persists regarding when and why his troubling finding applies. This paper provides a contemporary theoretical justification for Meehl's result. Importantly, Meehl's prediction problems require a small set of possible outcomes and machine-readable data. Second, individual predictions and decisions are evaluated only on average. This formulation leads to a natural analysis from statistical decision theory, which shows that statistical rules are more accurate than clinical intuition almost by definition. Meehl's prediction paradox is an example of \emph{metrical determinism}, where the rules of evaluation implicitly determine the best procedure. The decision-theoretic analysis of Meehl's problem elucidates the utility of algorithmic systems as decision-support tools, but also reveals their natural shortcomings, inducing expertise erosion, decision fatigue, and the usurpation of discretionary judgment.}

\begin{document}

\maketitle

\vspace{-16px}
\noindent{\small {\bf keywords:} {Statistical prediction. Decision theory. Metrical determinism.}}

\vspace{32px}

Throughout his undergraduate and graduate studies in Minnesota, Paul Meehl found himself at the center of a personal and professional conflict for the soul of psychology. As a high schooler, Meehl had been drawn to psychology by the psychodynamic school (pioneered by Freud), which considered the myriad connections between a patient’s past experiences–even their dreams–and their current mental state. At Minnesota, he was educated by a rigid behaviorist faculty, strongly anti-Freudian, focused on understanding the impact of external factors on mental states, and adamantly scientific and statistical. 

This struggle in psychology was part of a broader struggle in social science between the idiographic and the nomothetic. The idiographic focuses on the particulars, on the individual, trying to make sense of the singular and unverifiable. The nomothetic approach focuses on the general, aiming to determine laws and principles that explain categories with clear, measurable characteristics. The idiographic treats every case as unique. The nomothetic treats every case as a statistic. One of Meehl’s most famous works, \emph{Clinical versus Statistical Prediction} \citep{meehl1954clinical}, grew out of a lecture series from 1947 that probed the boundary between the purviews of the idiographic and the nomothetic in the human-facing sciences.

Although he wouldn’t call it by name, which wouldn’t be coined by Arthur Samuel for another half decade, Meehl made the first argument for machine learning in the clinic. After struggling to find a publisher, his book finally appeared in 1954, two years before the famous Dartmouth AI conference. It was four years before Rosenblatt’s Perceptron. Even as computers were just coming online, there was already ample evidence that statistical pattern recognition could, and perhaps should, play a role in critical decision making.

Meehl’s book focuses on predicting behavior. Some examples of the sorts of questions he wanted answers for include
\begin{itemize}
\item Given an application with LSAT score, undergraduate grades, and letters of recommendation, who should be admitted into law school?
\item Given a record of behavior, should a jailed person be released on parole?
\item Should you hospitalize a patient who is clinically depressed to prevent suicide?
\item Should a person who doesn’t respond to antidepressant prescriptions be given shock therapy?
\end{itemize}

These questions demand consequential decisions about people’s lives. They are all concerned with how a human being reacts under particular circumstances. And in all cases, the outcomes are uncertain. 

To answer such questions, we have to predict what will happen as a result of our actions. Such predictions are inherently probabilistic: We don’t know what will happen, so we must necessarily make uncertain assertions. The answers to the above questions necessarily take the form
\begin{itemize}
\item	“I believe this candidate will do well in our law program.”
\item “I believe this person will not commit crimes if released.”
\item “I believe this person will harm themselves unless they are committed.”
\item “I believe this person will find some relief from shock therapy.”
\end{itemize}
Meehl sought to determine the most effective way to quantify such beliefs. Moreover, given such quantification, what would be the best way to decide how to act upon an individual in the face of the inherent uncertainty of the future?

How can we estimate probabilities about the future? A nomothetic approach would be statistical, transmuting an assessment of past rates into future uncertainty. We could look at all similar cases in the past and count the number of times a treatment was successful. We could use the success percentage as a proxy for our belief that the treatment will work on this patient before us. Then, we could use optimal statistical decision rules to weigh the costs and benefits and select an action we believed would be most effective. This is statistical prediction. We convert past performance into future confidence. As discussed in detail by \citet {dawid_individual_2017}, the translation of past frequencies into future, individualized risk is so common that we seldom notice when this rhetorical turn is made in our faces.

Idiographic prediction, on the other hand, starts from the idea that all patients are unique. In the 1940s, the validity of statistical inference from class membership was not at all conventional wisdom. In his book, Meehl quotes Gordon Allport making the case for the idiographic: 
\begin{quote}
A fatal non-sequitur occurs in the reasoning that if 80\% of the delinquents who come from broken homes are recidivists, then this delinquent from a broken home has an 80\% chance of becoming a recidivist. The truth of the matter is that this delinquent has either 100\% certainty of becoming a repeater or 100\% certainty of going straight. If all the causes in his case were known, we could predict for him perfectly (barring environmental accidents). His chances are determined by the pattern of his life and not by the frequencies found in the population at large. Indeed, psychological causation is always personal and never actuarial. \citep{allport1942}
\end{quote}

Cases are indeed unique. Meehl himself concurs with Allport’s general sentiment.  But does that mean it is always impossible to make inferences from class membership? That seems too strong. Moreover, you’ll note that Allport uses the word “chances.” Chance is, by its very nature, a probabilistic concept. The question remains whether that chance can be usefully estimated through actuarial methods. When is generalizing about the future just a question of careful counting?

The question of when we can apply statistics feels quaint in our world obsessed with data and computation. We are living in the glory days of statistical pattern recognition. The tech industry and half of academia have decided that general intelligence is nothing but making decisions by counting things in appropriate reference classes. Everything we do is a sum of our past experience. All decisions are actuarial. It’s just a matter of finding the formula. But integrating statistical technique into the treatment of people remained an open and contentious question in the 1940s. Bradford Hill, one of the most influential epidemiologists of the 20th century and early advocate of the randomized clinical trial in medicine, made his case for statistical thinking as follows:

\begin{quote}
 “...the most frequent and the most foolish criticism of the statistical approach in medicine is…each patient is ‘unique’ and so there can be nothing for the statistician to count. But if this is true it has always seemed to me that the bottom falls out of the clinical approach as well as the statistical. If each patient is unique, how can a basis for treatment be found in the past observations of other patients?” \citep{hill_clinical_1952}
\end{quote}

Hill continued: “[Clinicians] base their ‘method of choice’ upon what they have seen happen before -whether it be in only two or three cases or in a score.”

Hill’s argument also has its own validity. When we were making decisions as clinical practitioners---whatever that may be, as teachers, as people on admissions committees, as people on tenure committees, or as physicians---every case that comes before us is read in reference to our evaluation of past cases. It is this grounding in experience that Hill argues is always statistical in nature.

Meehl would come down somewhere in the middle of Allport and Hill’s positions, though would end up closer to Hill. In 1947, before computers, before machine learning, and before AI, he attempted to understand the effectiveness and limitations of actuarial tables in human decision making. In what follows, I will walk through Meehl’s argument. I’ll make precise the sets of questions, decisions, and evaluation methods he considers. I’ll provide his evidence. And I’ll close with some reflections on the valuable lessons we can still learn from reading Meehl’s 1954 book.

\section*{Meehl's Formalized Decision Problems}

Where do we draw the line between where statistics applies and where it doesn’t? If you are in a casino, and you trust the house to play fair, we’d probably all agree that the outcomes of future card games can be statistically analyzed. When creating actuarial tables to price insurance, the risks and prices are all based on carefully computed relative frequencies. The insurance company has found this mindset useful enough to build a business on top of it for centuries. But if a doctor is operating on a patient with an extremely uncommon condition, is that statistics too? In a sense, we can only define the term 'uncommon' in statistical terms. It refers to a relative frequency of occurrence. However, in these cases where experiences do seem wholly new, how can we map past rates onto how to act? There is clearly a spectrum between when pure statistics can guide action (e.g., betting on blackjack) and where perhaps there is something else that must be applied (e.g., surgery on a novel condition).

To get after the merits between the clinical and statistical, Meehl strongly boxes in the scope of questions to be answered. One of the central aims was to determine the \emph{scope} of utility of statistical judgment. There was a significant set of decisions where he deemed statistics superior. By being precise about this subset, he thought that he could both improve care and simplify the life of the clinician, allowing them room to automate part of their job. 

\paragraph{Actions.} Meehl first clarifies that the goal should be about predicting the outcome of interventions. All of the example questions he asks are attempting to predict how an action will affect a particular person. If granted admission, will a person succeed in law school? If released from prison, will a person recidivate? If a depressed person isn’t hospitalized, will they commit suicide? If a person receives shock therapy, will their depression be relieved? 

These types of questions concern the impact of individual actions. They also have yes-or-no answers. Meehl focuses on questions with a small list of possible outcomes. For open-ended questions, Meehl thought clinical expertise was indispensable. It was only for problems with simple multiple-choice answers where he thought statistical decision making could play a role.

\paragraph{Data.} To make the decision, Meehl assumes the clinician has the same data as the statistical rule. He belabors the distinction between the kind of data and the mode of combining the data. As long as the statistical formula and the clinician receive the same information, the data can be anything, be it interviews, life history, mental tests, or other biometrics. 

Obviously, such data has to be transformed into a machine-readable format somehow. Here’s another place where the clinician may be indispensable. A clinician may be required to observe a patient’s behavior or facial expressions and record the appropriate diagnostic information. Machine-readable certainly meant something very different in 1954 than in 2025. In his reflections, \citet{meehl_causes_1986} notes that character recognition is still barely functional. He doesn’t rule out the possibility of more sophisticated pattern recognition methods emerging if computers improve. Of course, they did. Now, patient data is charted in detailed electronic health records, and these can be processed by large language models. The definition of machine-readable is growing larger every day.
\paragraph{Mechanical and actuarial rules.} Meehl defines two forms of algorithmic decision rules. First, there are mechanical rules, which we now refer to as algorithms. Mechanical rule and algorithm are synonymous. A mechanical rule is a well-defined, step-by-step process for translating data into a decision that can be implemented on a computer. 

Actuarial rules are a special kind of mechanical rule. They are algorithms that make decisions based on rates of past occurrences. These are the statistical prediction methods. A decade ago, we called these prediction methods machine learning. Today, we call them AI.

Many mechanical rules are not pure statistical prediction. For example, whether or not you qualify for a tax credit is based on a hard threshold based on certain attributes of your income and expenses. Actuarial prediction thus takes the following form: we identify the best attributes to represent our data in a machine-readable format. We build actuarial tables of this data. Importantly, we assert that we believe the future will be like the past. Based on this assertion, we maximize our outcomes accordingly. 

\paragraph{Clinical judgment.} Meehl’s definition of a clinical judgment is deliberately more vague. He says it’s anything “informal” made by a human specialist. It’s whatever process occurs in a person’s head. Clinical rules are those made by clinicians based on intuitive assessments of data. These are decisions that clinicians can’t cleanly explain and hence may not be easily formalized as mechanical algorithms. 

\paragraph{The clinical-statistical question.}

With all of this setup, we can now pose Meehl’s central question:
\begin{quote}
Given a decision problem with a small set of possible outcomes and an appropriate, fixed collection of data, do actuarial rules or clinical judgment provide more accurate judgments about the future?
\end{quote}
For this narrow but broadly applicable question, Meehl came down solidly on one side: Statistical prediction would never be worse than clinical prediction.

Meehl provides compelling empirical evidence in his 1954 book. And 70 years of studies have backed him up. You’d be hard-pressed to find a result in social science that is as robust as statistical decisions outperforming clinical judgment.

I will now present both the empirical evidence, Meehl’s philosophical arguments, and what I consider to be a simple but deceptively subtle explanation. It’s through the subtlety that we might find some resolution.

\section*{A Century of Evidence}

The earliest study Meehl finds demonstrating the superiority of statistical judgment asked whether “scientific methods” could be applied to parole. In the 1920s, sociologist Ernest Burgess worked with the Illinois Parole Board to determine the factors that contributed to recidivism and whether it was possible to predict whether a parolee would commit further crimes after release. 

\citet{Burgess1928} assembled 21 predictive factors, including age, the type of offense, whether a person was a repeat offender, and whether the person had held a job before. He then constructed a sophisticated AI tool for predicting parole: he scored each factor either 0 or 1 and then added them all up. Of the 68 men with at least 16 positive factors, only one ever committed a crime again. Of the 25 men with fewer than five positive factors, 19 recidivated.

In a 1928 report, Burgess compared his predictions to those of two prison psychiatrists. He gathered a dataset of 1000 men who appeared before the Illinois Parole Board in the 1920s. The psychiatrists assigned each prisoner as likely to violate parole, unlikely to violate parole, or uncertain. Of the ones deemed unlikely to violate, the first psychiatrist predicted 85 percent correctly, the second 80 percent. Of the individuals deemed likely to violate parole, the first psychiatrist predicted 30 percent correctly, while the second predicted 51 percent. Burgess’ method, looking for at least ten positive factors, not only made a prediction for all parolees but also correctly predicted 86 percent of those unlikely to violate and 51 percent of those likely to violate. It outperformed the first psychiatrist at predicting violations and the second at predicting successful parole.

Algorithmic recidivism prediction remains a contentious topic. It is one of the most frequently discussed examples of human-facing prediction in the machine learning fairness community, where it is often cast in a negative light. For example, \citet{FraenkelBook} argues recidivism scores are “an opaque decision-making system that influences the fundamental rights of residents of the US.” \citet{Wang24} say ``formulating pre-trial risk as a prediction problem does not do anything to increase our understanding of the underlying phenomenon, nor does it help us discover better interventions.'' And yet, Burgess was attempting to make the case for a more liberal parole system. He thought his algorithm could be less political, more fair, and more accurate. Perhaps it would be more just to remove the discretion of biased judges.

Meehl highlights a dozen other studies in his book and continued to track examples throughout his career. No matter how much he looked, he kept finding the same thing: statistical rules were seldom worse and often much better than clinical predictions. In a reflection on his book, Meehl wrote in 1986, “There is no controversy in social science that shows such a large body of qualitatively diverse studies coming out so uniformly in the same direction as this one.” 

Two subsequent meta-analyses were done by \citet{grove_clinical_2000} and \citet{aegisdottir_meta-analysis_2006}. Grove et al.’s analysis included 136 predictions. In 46\% of the predictions, mechanical methods were roughly five percentage points better than clinical judgments. That is, the difference in accuracy between the statistical and clinical predictions was at least 0.05. In 48\%, the predictions were within about 5 points of each other. Clinical predictions were substantially better than mechanical predictions in less than 6\% of the studies. Moreover, \citet{grove_clinical_2000} found a substantial skew in the distribution: When they were better, mechanical predictions were more frequently far better than their clinical counterparts.

{\AE}gistd\'{o}ttir et al. focused on statistical methods but found the same results as Grove et al. In their compilation of 48 predictions, 52\% favored statistical methods, 38\% reported comparable performance, and 10\% favored clinical judgment.

What can we make of these results? Many feel like a doctor can assess more than what is fed into the computer. That a counselor can see subtle cues that are valuable for prediction. That there are edge cases that statistical algorithms can’t catch. Why does the empirical evidence not bear this out? Why does clinical judgment repeatedly fare worse on average?

The key to the entire clinical-statistical puzzle is those last two words. 

\section*{A Rigged Game}

The trick that Meehl plays is in the quantification of “better.” By better, we of course mean on average. This is a subtle point: Meehl discusses in Chapter 4 that a clinician may be able to detect a variety of exceptional cases that don't appear in the original data seen by the statistical algorithm. His famous example is where an actuarial table determines that Professor Glotz attends the movies 90\% of all Fridays, but this Friday he has a broken leg. The broken leg impels the clinician to change their predicted probability to near zero. What if clinicians are adept at finding such idiographic oddities as broken legs? Meehl doesn't deny this possibility, but asserts that, regardless of how clinicians incorporate new knowledge, their performance should be evaluated actuarially. As Meehl puts it:

\begin{quote}
The clinician may be led,... to a guess which turns out to be correct because his brain is capable of that special "noticing the unusual" and "isolating the pattern" which is at present not characteristic of the traditional statistical techniques. Once he has been led to a formulable sort of guess, we can check up on him actuarially.
\end{quote}

Actuarial evaluation seems innocuous: how else would we compare two decision-makers but by the body of their work? However, once all parties decide that predictions will be evaluated by averages, the game is up. If prediction is possible, meaning that the past and the future are similar, and the evaluation is based on rates of future success, then the best predictor will be the one that maximizes success rate among some class of possible algorithms. You should find a rule that accurately predicts the past and use it to make predictions about the future. Since you will be evaluated based on averages, this is effectively the optimal thing to do. 

We can make this heuristic argument precise. In what follows, I follow a standard argument from decision theory invented by  \citet{wald_statistical_1945}. A more modern treatment can be found in Detection Theory textbooks, such as \citet{KayDetectionBook}. These arguments form the foundation for the ubiquitous method of \emph{empirical risk minimization} in machine learning (see, for example, the development in \citet{DudaHartBook}). I'll belabor the development to highlight the connections to Meehl's restricted form of decision problems.

 The argument proceeds by stratifying populations into groups defined by the data and computing the optimal decision rule for each stratum. The optimal prediction is thus always a statistical function of each stratum. In the language of probability, it reveals that predictions depend only upon the conditional probability of an outcome given the data used to make a decision. 
 
We are interested in making predictions about a set, $I$, of individuals. For each individual $i \in I$, a method must assign a prediction $p_i$. Each prediction must be made based only on a set of machine-readable data, denoted by $x_i$. Adopting language from machine learning, we will call the vector of assembled data the \emph{features} used for prediction. Each individual experiences some outcome $y_i$ that we are trying to predict from clinical measurement. For simplicity, assume this outcome is binary valued, taking one of two values $0$ or $1$. The generalization to multiple outcomes will not lead to different conclusions.

Each prediction is quantitatively scored by a scoring function $S(p,y)$. This is a metric decided in advance. This score could be the amount of cost accrued for predicting $p$ when an individual had propensity $y$. It could be simply an indicator of accuracy, where $S(p,y)$ is equal to $1$ when $p=y$ and $0$ otherwise. Or it could score probabilistic predictions where $p$ is a probabilistic prediction taking a value between $0$ and $1$. 

In all of these cases, the performance of a forecaster is evaluated by its average score on $N$ individuals.
\begin{align*}
S_{\text{avg}} = \frac{1}{N} \sum_{i \in I} S(p_i,y_i) 
\end{align*}

Once a judge observes all of the outcomes, they can retrospectively compute a well-defined \emph{optimal} prediction. It is the prediction that minimizes the above expression. We can compute the optimal decision rule by rearranging terms in this summation. 

Note that under Meehl's rules, all individuals with the same feature vector $x$ should be assigned the same prediction. That is, if $x_i=x_j$, then we must have $p_i=p_j$. With this in mind, we can collect particular statistics about each value of the feature vector. Let $q_x$ denote the fraction of the time that $y_i=1$ when $x_i=x$. The reader versed in probability can think of the quantity $q_x$ as the conditional probability of $y$ given $x$ under the uniform distribution on the discrete set $\{(x_i,y_i)\}_{i\in I}$. Also, let $n_x$ denote the number of individuals in the data set with $x_i=x$. Let $\mathcal{X}$ denote the set of all possible values of the feature vector $x$. With this notation, we can collect terms in the average score to reveal the optimal predictions.
\begin{align*}
S_{\text{avg}} &= \frac{1}{N} \sum_{i \in I} S(p(x_i),y_i) \\
&=\sum_{x \in \mathcal{X}} \frac{n_x}{N} \left\{\frac{1}{n_x} \sum_{\{i~:~x_i=x\}} S(p(x),y_i)\right\}\\
&= \sum_{x \in \mathcal{X}} \frac{n_x}{N} \left\{ q_x S(p(x),1) + (1-q_x) S(p(x),0) \right\}
\end{align*}
Once the individuals are stratified, the optimal predictions are deterministic functions of the data-conditional rates, $q_x$. That is, if we knew all of the outcomes in advance, the optimal prediction in retrospect is a function only of the rate of positive events of those individuals with the same valued data. For an individual presenting with data $x$, the optimal prediction minimizes the inner expression in the summation
$$
	p^\star(x) \in \arg \min_ p  q_x S(p,1) + (1-q_x) S(p,0)\,.
$$
Given the data, the only information we need to know is the rate of outcomes of individuals with the same data and the metric under which the predictions will be evaluated. Regardless of the score function $S$, the optimal prediction is inherently statistical. It is only a function of the set of individuals who are grouped as identical under the agreed-upon data format. 

As an illustrative example, let's consider the simple case when $S$ is the \emph{Brier Score}, a popular metric used to evaluate predictions. In this case, $S(p,y)=(p-y)^2$. For the Brier score, an elementary calculation reveals the optimal prediction $p^\star(x)=q_x$. That is, the optimal prediction when forecasting under the Brier score is precisely the outcome rate for that particular group. The Brier Score is an example of a \emph{proper scoring rule}, and it is also true that $p^\star(x)=q_x$ for any such scoring rule. If a forecaster is evaluated by a proper scoring rule, their optimal predictions are the average outcomes for every group. For more on proper scoring rules and their properties, I refer the reader to \citet{gneiting_strictly_2007}. 

Of course, for these evaluations, we don't know the rates $q_x$ in advance. However, if we could accurately predict these rates, we would optimally solve the prediction problem. A reasonable guess for rates $q_x$ on individuals we haven't seen is the rates on the individuals we \emph{have} seen. If we collect actuarial tables and make predictions based on such rates, we can expect good prediction scores when the individuals on whom we will be evaluated are similar to those we have already seen. If you had any indication that these rates would change for the new collection of individuals, Meehl's problem formulation allows the statistical account to accommodate this. The same information must be given to the statistical processor as to the clinician. Furthermore, if the new individuals have outcomes completely different from past experience, then both the clinician and the actuary are stuck.

If I pick a statistical evaluation, I can derive the optimal decision. I call this phenomenon, where the metric fixes optimal actions, \emph{Metrical Determinism}. The evaluation ties our hands. Once we decide what is best in the future, the problem of optimal action is mechanical. It should thus not be surprising that statistics wins when we evaluate predictions and decisions using statistics.

Meehl summarizes the situation in the last paragraph of his 1954 book. If we subscribe to the bureaucratic utilitarian mindset, the algorithm always wins:
\begin{quote}
“If a clinician says, ‘This one is different’ or ‘It’s not like the ones in your table,’ ‘This time I’m surer,’ the obvious question is, ‘Why should we care whether you think this one is different or whether you are surer?’ Again, there is only one rational reply to such a question. We have now to study the success frequency of the clinician’s guesses when he asserts that he feels this way. If we have already done so and found him still behind the hit frequency of the table, we would be well advised to ignore him. Always, we might as well face it, the shadow of the statistician hovers in the background; always the actuary will have the final word.”
\end{quote}

\section*{When Shall We Use Our Heads Instead of the Formula?}

Given the dark and heated rhetoric of Meehl's closing, it isn't surprising his book has been met by the persistent, angry disbelief of clinicians. This contentiousness, unfortunately, meant the central results were often misunderstood. One of the more common misreadings of Meehl claims he was arguing to do away with clinicians altogether. This was not his position at all, as he highlights in early follow-up commentary \citep{meehl_when_1957}  and reiterates in reflections thirty years later \citep{meehl_causes_1986}.  Meehl did not believe that all decisions could be made statistically. 

In fact, there is a far more positive spin on his analysis. Meehl provides clinicians with clearly delineated conditions for when statistical methods are useful: answering clear, multiple-choice questions about simple actions from machine-readable data. This characterization is useful in of itself.

Moreover, I cannot emphasize enough here that just because statistical prediction is never worse and often better than clinical judgment, that doesn’t mean that it isn't possible to poorly implement statistical prediction. Careful statistical prediction remains a delicate skill. You can have too few features to make accurate predictions. You can have too many features, making it hard to find consistent patterns. You might be in a situation where you have completely uninformative features. We don't have particularly effective methods to deal with missing data, and missing data plagues many prediction problems about people. 

Most worrisomely, the predictions trained on statistical counts have limited temporal validity, as the population of people changes faster than the statistical prediction rules can be updated. Statistical prediction relies on past counts being reasonable predictions of the future. We have plenty of experience that tells us this is often not a safe assumption. In practice, statistical rules quickly lose predictive accuracy \citep{Vela2022,munger2023temporal} and require massive effort to keep them updated. Data scientists and software engineers at technology companies refer to this degradation as \emph{staleness} (see, for example, \citep{koren2009collaborative}) and constantly retrain their prediction systems to prevent predictions from becoming less accurate. Not all fields are as diligent about the maintenance of their prediction systems. Medical risk assessments may remain static for decades, although they become ineffective within a matter of years \citep{garcia2014validity}.

These are just a few of the major headaches one has to deal with when deploying statistical decision aids. If we’re going to rely on statistical prediction, then we need expertise in statistical hygiene.

Beyond these statistical issues, widespread automated decision making entails high human costs. Too many mechanical rules can lead to an erosion of expertise as practitioners spend too much time deferring to their apps. They can lead to decision fatigue, forcing too many things into overwhelming, computerized systems. And they can lead to complacency, as following mechanical rules is drudgery. (\citet{KleinSnSBook} has a thorough discussion of these costs). For all these reasons, the adoption of mechanical rules and statistical prediction in high-stakes scenarios must be done with care.

Moreover, statistical rules need to be targeted at interventions with simple outcomes. Trying to shoehorn every decision into a simple statistical decision narrows the possibilities of the world we inhabit. The Meehlian actuarial game transforms the world into machine language. This is explicitly part of the problem setup, which demands machine-readable rules, data, and outcomes. The game is rigged because we organized the problem to be mechanical. Once the problem is mechanical, it can be solved by a machine. However, if machines can’t function, they have no role in decision making. We can only compare human to machine decisions on the problems where we level the playing field for the machine.

Nonetheless, one of the primary impulses of the modern state is to translate human experience into data readable by machines.  Bureaucracies render humanity in a simplified state in order to make decisions about it. And, as explicated by \citet{FarrellFourcade}, our massive technology companies aid, abet, and profit from helping with such rendering. These systems remove the discretion of people in the decision making chain. These people, be they your primary care physician or a trial judge, often consider benefits not captured in actuarial evaluations. For those impacted by a decision, the impact on the individual is more important than the impact on the average. Those of us who aren't bureaucrats should be careful about which discretion we remove from decisions that affect people's lives. We should tell the actuary that their word isn't the end after all.

{\small
\bibliography{actuarial}
\bibliographystyle{plainnat}
}
\end{document}